\documentclass[twoside, a4paper]{llncs}

\usepackage[utf8]{inputenc}
\usepackage[T1]{fontenc}
\usepackage{./why3lang}
\usepackage{amsmath}
\usepackage{cite}
\usepackage{xspace}

\newcommand{\why}{\textsf{Why3}\xspace}

\usepackage{hyperref}

\title{Desfuncionalizar para Provar}
\author{Mário Pereira\thanks{Este trabalho foi realizado enquanto o autor se
    encontrava afiliado ao LRI -- Laboratoire de Recherche en Informatique,
    CNRS, Université Paris-Sud -- na qualidade de aluno de doutoramento.
  }}

\institute{NOVA -- LINCS}

\begin{document}
\maketitle

\pagestyle{plain}

\begin{abstract}
  Este artigo explora a ideia de utilizar a desfuncionalização como uma técnica
  de prova de programas de ordem superior. A desfuncionalização consiste em
  representar valores funcionais por uma representação de primeira. O seu
  interesse é portanto poder utilizar posteriormente uma ferramenta de prova de
  programas existente, sem ser necessário estender esta ferramenta com suporte
  para ordem superior. Este artigo ilustra e discute esta abordagem através de
  diversos exemplos práticos, construídos e validados na ferramenta de
  verificação dedutiva \why{}.
\end{abstract}

\section{Introdução}
\label{sec:introducao}

Um programa de ordem superior é um programa que utiliza funções como valores de
primeira classe. Em tais programas, funções podem ser passadas como argumentos a
outras funções ou devolvidas como o resultado de computações. O conceito de
ordem superior é amplamente conhecido e utilizado em linguagens ditas
\emph{funcionais}, tais como OCaml, Haskell ou SML. Recentemente, linguagens
como Java ou C++ introduziram suporte para funções de ordem superior.

A prova de correcção de programas de ordem superior coloca desafios complexos,
em particular no contexto de programas com efeitos. Metodologias
existentes~\cite{nanevski08icfp,chargueraud11icfp} empregam assistentes de prova
interactivos e uma codificação dos efeitos directamente na lógica destes
assistentes de prova. No contexto de prova automática de programas, Kanig e
Filliâtre~\cite{KanigFilliatre09wml} propuseram um sistema de tipos e efeito,
assim como um cálculo de pré-condições mais fracas, com o propósito de
especificar e provar a correção funcional de programas de ordem superior. No
entanto, tal sistema é difícil de utilizar de forma prática. 

Neste artigo, propomos uma nova metodologia para a verificação de programas de
ordem superior que potencialmente comportam efeitos secundários. Exploramos a
ideia de empregar a técnica de desfuncionalização para verificar tais
programas. A utilização da desfuncionalização permite-nos obter um programa de
primeira ordem equivalente ao programa de ordem superior original, o que nos
conduz a um contexto de programas de primeira ordem. Assim, podemos recorrer a
ferramentas de prova existentes, sem que para tal seja necessário estender tais
ferramentas com suporte de ordem superior. Esta técnica apresenta como limitação
o facto de ser necessário conhecer \textit{à priori} todas as funções que serão
utilizadas como valores de primeira classe. Mesmo se, teoricamente, tal nos
impede de aplicar a técnica de desfuncionalização para provar qualquer programa,
é possível identificar uma classe interessante e representativa de programas de
ordem superior aos quais nos é possível aplicar a desfuncionalização para provar
a sua correcção. Neste artigo, apresentamos a nossa abordagem através de
diversos exemplos escritos e verificados no sistema de prova de programas
\why{}~\cite{filliatre13esop}.





Este artigo encontra-se organizado como se segue. A
secção~\ref{sec:desfuncionalizacao} introduz a técnica de desfuncionalização,
com base num exemplo completo. Na secção~\ref{sec:preuve-par-defonct}
apresentamos em detalhe a nossa abordagem de utilização da desfuncionalização
como meio de prova. Ilustramos a nossa proposta através de diferentes
exemplos. O artigo termina sobre algumas perspectivas e conclusões sobre o nosso
trabalho. O código \why\ apresentado neste artigo pode ser encontrado no
seguinte endereço electrónico: \url{http://www.lri.fr/~mpereira/defunc}.

\section{Desfuncionalização}
\label{sec:desfuncionalizacao}

A desfuncionalização é uma técnica de transformação de programas de ordem
superior para programas de primeira ordem. Esta técnica foi introduzida por
Reynolds~\cite{reynolds-98a} como forma de obter um interpretador de ordem
superior a partir de um interpretador de primeira ordem. 
Recentemente, esta técnica tem sido amplamente explorada por Danvy e seus
colaboradores~\cite{danvy01ppdp,danvy09scp}. Em particular, estes autores
mostraram de que forma é possível derivar máquinas abstractas para diferentes
estratégias de avaliação do cálculo-lambda a partir de interpretadores
composicionais~\cite{DBLP:conf/ppdp/AgerBDM03}.

Passamos a explicar o princípio da desfuncionalização através de um exemplo
escrito em OCaml\footnote{Este código pode ser facilmente adaptado a qualquer
  linguagem que suporte funções como valores de primeira classe.}
Consideremos o programa que calcula a altura de uma árvore binária, escrito em
estilo CPS~\cite{Plotkin75}:
\begin{why3}
  type 'a tree = E | N of 'a tree * 'a * 'a tree

  let rec heigth_tree_cps (t: 'a tree) (k: int -> 'b) : 'b = match t with
    | E -> k 0
    | N (l, x, r) ->
       heigth_tree_cps l (fun hl ->
       heigth_tree_cps r (fun hr -> k (1 + max hl hr)))

  let heigth_tree t = heigth_tree_cps t (fun x -> x)
\end{why3}
Este programa foi cuidadosamente escrito em estilo CPS para evitar quaisquer
problemas de \textit{stack overflow}, independentemente da forma da árvore.
A transformação CPS pode ser usada como um meio automático para evitar problemas
de \textit{stack overflow}, desde que o compilador optimize chamadas recursivas
terminais.

Chamamos a atenção para a presença, na função \of{heigth_tree_cps}, do argumento
\of{k} do tipo \of{int -> 'b}. É a utilização deste argumento que confere à
função \of{heigth_tree_cps} a sua natureza de ordem superior. Este argumento
actua como uma \emph{continuação}, ou seja, ao invés de devolver um resultado, a
função \of{heigth_tree_cps} passa-o como argumento a~\of{k}. No caso da árvore
vazia (primeiro ramo da filtragem), por exemplo, aplicamos~\of{k} a 0 em
detrimento de devolver directamente tal resultado.

Três funções anónimas são utilizadas no código do programa que calcula a altura
da árvore. Na chamada \of{heigth_tree_cps l} a continuação \of{(fun hl -> ...)}
é aplicada ao resultado do cálculo da altura da sub-árvore esquerda~\of{l}. O
argumento~\of{hl} representa a altura de~\of{l} já calculada. No interior desta
continuação encontramos uma segunda chamada recursiva com uma outra continuação,
desta vez sobre a sub-árvore direita~\of{r}. Aquando da aplicação da função
anónima \of{(fun hr -> ...)}, conhecemos já todos os ingredientes necessários
para calcular a altura da árvore inicial~\of{t}. Também neste caso não
devolvemos directamente o resultado da expressão \of{1 + max hl hr}, mas é sim
passado a~\of{k}. A terceira e última função anónima que encontramos neste
programa é a função identidade \of{(fun x -> x)}. A aplicação desta função
garante que o resultado devolvido por \of{heigth_tree t} é de facto a altura da
árvore~\of{t}.

Para desfuncionalizar este programa, precisamos de uma representação de primeira
ordem para as três abstracções utilizadas. Para isso, substituímos o tipo
funcional por um tipo algébrico, através do qual capturamos os valores das
variáveis livres utilizadas em cada função. Para o exemplo apresentado acima,
introduzimos o tipo seguinte:
\begin{why3}
  type 'a kont = Kid | Kleft of 'a tree * 'a | Kright of 'a kont * int
\end{why3}
O construtor \of{Kid} representa a função identidade e por isso não contém
variáveis livres. Os construtores \of{Kleft} e \of{Kright} representam,
respectivamente, as funções \of{(fun hl -> ...)} e \of{(fun hr -> ...)}. Os seus
argumentos capturam as variáveis livres utilizadas em cada uma das duas
funções. No caso de \of{Kleft} guardamos a sub-árvore \of{r} e a continuação
\of{k}; no caso de \of{Kright} o primeiro argumento é o valor de~\of{k} e o
segundo representa~\of{hl}, a altura da sub-árvore esquerda.

\begin{figure}[t]
  \centering
  \begin{why3}
  type 'a tree = E | N of 'a tree * 'a * 'a tree

  type 'a kont = Kid | Kleft of 'a tree * 'a kont | Kright of 'a kont * int

  let rec apply (k: 'a kont) arg = match k with
    | Kid -> let x = arg in x
    | Kleft (r, k) -> let hl = arg in heigth_tree_cps r (Kright (k, hl))
    | Kright (k, hl) -> let hr = arg in apply k (1 + max hl hr)
  and heigth_tree_cps t (k: 'a kont) = match t with
    | E -> apply k 0
    | N (l, x, r) -> heigth_tree_cps l (Kleft (r, k))

  let heigth_tree t = heigth_tree_cps t Kid
  \end{why3}
  \caption{Programa que calcula a altura da árvore, desfuncionalizado.}
  \label{fig:heigth_tree_cps_def}
\end{figure}

Tendo entre mãos uma representação de primeira ordem, podemos agora proceder à
substituição de todas as abstracções pelo construtor de \of{'a kont}
correspondente:
\begin{why3}
  let rec heigth_tree_cps t (k: 'a kont) = match t with
    | E -> ??? (* aplicar k *?\textit{desfuncionalizado}?* ao seu argumento *)
    | N (l, x, r) -> heigth_tree_cps l (Kleft (r, k))

  let heigth_tree t = heigth_tree_cps t Kid
\end{why3}
A segunda etapa do processo de desfuncionalização consiste em introduzir uma
função \of{apply} para substituir as aplicações no programa original. Esta
função aceita como argumentos um valor do tipo \of{kont 'a} sobre o qual é
realizada uma análise por casos. Para cada construtor, a função \of{apply}
executa o código da abstracção que corresponde a esse mesmo construtor. O
segundo argumento de \of{apply} trata-se do argumento da aplicação no programa
original. A função \of{apply} para o nosso exemplo é a seguinte:
\begin{why3}
  let rec apply (k: 'a kont) arg = match k with
    | Kid -> let x = arg in x
    | Kleft (r, k) -> let hl = arg in heigth_tree_cps r (Kright (k, hl))
    | Kright (k, hl) -> let hr = arg in apply k (1 + max hl hr)
\end{why3}
Para obter o programa completamente desfuncionalizado, basta então substituir
todas as aplicações da continuação \of{k} por chamadas à função \of{apply}. O
código completo resultante para este exemplo pode ser encontrado na
Fig.~\ref{fig:heigth_tree_cps_def}.





\section{Prova por desfuncionalização}
\label{sec:preuve-par-defonct}

Nesta secção exploramos a ideia de utilizar a desfuncionalização como uma
técnica de prova para programas de ordem superior com efeitos. A nossa proposta
é a seguinte:
\begin{enumerate}
\item dado um programa de ordem superior (possivelmente contendo efeitos),
  juntamos-lhe uma especificação lógica.
\item 
  desfuncionalizamos este programa e ao mesmo tempo traduzimos a sua
  especificação a fim que esta se torne uma especificação do programa
  desfuncionalizado.
\item utilizamos uma ferramenta de prova de programas existente para produzir
  a prova de que o programa desfuncionalizado respeita a especificação dada.
\end{enumerate}
Passamos a ilustrar esta proposição através de diferentes exemplos: o programa
que calcula a altura de uma árvore (Sec.~\ref{sec:hauteur-dun-arbre}); um
programa que calcula o número de elementos distintos de uma árvore
(Sec.~\ref{sec:nombre-delem-diff}); um interpretador \textit{small-step} para
uma pequena linguagem (Sec.~\ref{sec:interpr-petits-pas}).  Todas as
experiências descritas são realizadas com ajuda da ferramenta de verificação
\why{}. Como actualmente o \why\ não nos permite racionar sobre programas de
ordem superior, em geral, todos os exemplos de especificação de programas de
ordem superior são escritos num sistema hipotético, uma possível extensão do
\why{}.




\subsection{Altura de uma árvore}
\label{sec:hauteur-dun-arbre}

Recuperemos o exemplo da Sec.~\ref{sec:desfuncionalizacao}. A fim de especificar
este programa, devemos instrumentar as funções \of{heigth_tree_cps} e
\of{heigth_tree} com contratos. A função \of{heigth_tree} devolve a altura da
árvore~\of{t} passada como argumento, tal como especificamos na sua
pós-condição:
\begin{why3}
  let heigth_tree (t: tree 'a) : int
    ensures { result = height t }
  = heigth_tree_cps t (fun x -> x)
\end{why3}
Aqui, \of{result} trata-se uma palavra-chave do \why\ para representar o valor
devolvido e \of{height} é uma função lógica que devolve a altura de uma árvore.

O valor devolvido pela função \of{height_tree_cps} é o resultado da aplicação da
continuação~\of{k} à altura da árvore~\of{t}. Seria assim natural de equipar
\of{height_tree_cps} com a seguinte pós-condição:
\begin{why3}
    ensures { result = k (height t) }
\end{why3}
No entanto, esta especificação coloca-nos o problema de como interpretar a
utilização de funções de programa na lógica. Em geral, tal utilização pode
facilmente produzir incoerências lógicas, já que funções de programa podem
conter efeitos, \emph{e.g.}, divergência. Fazemos então a escolha de restringir
a nossa linguagem de especificação: podemos utilizar os nomes de funções de
programa, mas impomos uma barreira de abstracção entre o mundo lógico e o
programa. Para esse fim, adoptamos um sistema no qual as funções são abstraídas
sob a forma de um par de predicados que representam as suas pré-condição e
pós-condição~\cite{regis-gianas-pottier-08}. Assim, no interior de uma fórmula
lógica, uma função~$f$ do tipo $\tau_1\rightarrow\tau_2$ é vista como o par de
predicados
\[
  \begin{array}{rl}
    \mathtt{pre} \: f : & \tau_1 \rightarrow \mathtt{prop} \\
    \mathtt{post} \: f : & \tau_1 \rightarrow \tau_2 \rightarrow \mathtt{prop}
  \end{array}
\]
Utilizamos $\mathtt{pre}\: f$ e $\mathtt{post}\: f$ para nos referirmos à
pré-condição e à pós-condição de~$f$, respectivamente.


Voltemos ao exemplo da altura de uma árvore. Podemos então escrever, utilizando
a notação apresentada acima, o contrato seguinte para a função
\of{height_tree_cps}:
\begin{why3}
  let rec height_tree_cps (t: tree 'a) (k: int -> 'b) : 'b
    ensures { post k (height t) result }
\end{why3}
Esta pós-condição impõe uma relação entre o valor passado a~\of{k} (a altura
de~\of{t}) e o resultado devolvido (\of{result}). Seguindo a nossa metodologia,
podemos igualmente fornecer contratos às funções anónimas utilizadas no interior
de \of{height_tree_cps}:
\begin{why3}
  let rec heigth_tree_cps (t: tree 'a) (k: int -> 'b) : 'b
    ensures { post k (heigth t) result }
  = match t with
    | Empty -> k 0
    | Node l x r ->
       heigth_tree l (fun hl -> ensures { post k (1 + max hl (height r)) result }
       heigth_tree r (fun hr -> ensures { post k (1 + max hl hr) result }
         k (1 + max hl hr)))
    end
\end{why3}
Para a primeira abstracção, especificamos que o resultado da sua aplicação se
relaciona com a altura da árvore completa, através da altura \of{hl} da
sub-árvore esquerda e da altura da sub-árvore direita, que ainda não
conhecemos. Juntamos à segunda abstracção uma pós-condição semelhante, com a
nuance de que no momento de aplicar esta função já conhecemos a altura \of{hr}.

Procedemos agora à desfuncionalização deste programa, assim como da sua
especificação, para em seguida provar a sua correcção funcional com a ajuda da
ferramenta \why{}. O código obtido é o seguinte:
\begin{why3}
  type kont 'a = Kid | Kleft (tree 'a) (kont 'a) | Kright int (kont 'a)

  let rec heigth_tree_cps (t: tree 'a) (k: kont 'a) : int
    ensures { post k (height t) result }
  = match t with
    | Empty -> apply k 0
    | Node l _ r -> heigth_tree_cps l (Kleft r k)
    end
  with apply (k: kont 'a) (arg: int) : int
    ensures  { post k arg result }
  = match k with
    | Kid -> arg
    | Kleft r k -> heigth_tree_cps r (Kright arg k)
    | Kright hl k -> apply k (1 + max hl arg)
    end

  let height_tree (t: tree 'a) : int
    ensures { result = height t }
  = heigth_tree_cps t Kid
\end{why3}
Como podemos observar, este programa é em tudo semelhante ao programa OCaml da
Fig.~\ref{fig:heigth_tree_cps_def}, excepção feita às pequenas diferenças
sintácticas em relação à linguagem do \why{} e à presença das anotações
lógicas. De notar que as funções \of{heigth_tree_cps} e \of{apply} se mantém
como funções mutuamente recursivas.

A especificação do programa desfuncionalizado é a mesma do programa original. Em
particular, mantemos a nossa utilização da projecção~\of{post} para a
pós-condição de \of{heigth_tree_cps}. Precisamos então de fornecer uma
especificação à função~\of{apply}, a nova função gerada pelo processo de
desfuncionalização. Como a função~\of{apply} simula a aplicação de uma função ao
seu argumento, a única especificação que lhe podemos fornecer é a de que a sua
pós-condição é a pós-condição da função~\of{k}. Da mesma forma, a pré-condição
de \of{apply} seria a pré-condição de~\of{k}, à qual podemos aceder utilizando a
projecção~\of{pre}. Escolhemos não o fazer para este exemplo, visto tratar-se da
pré-condição trivial.

Finalmente, precisamos de introduzir o predicado \of{post} na especificação
lógica desfuncionalizada. Para tal, criamos um predicado \of{post} que reúne as
pós-condições fornecidas no programa original. Tal como para a função
\of{apply}, um tal predicado efectua uma filtragem sobre o tipo algébrico
\of{kont 'a} e para cada construtor copiamos a pós-condição fornecida na
abstracção correspondente. Para este exemplo, o predicado \of{post} é o
seguinte:
\begin{why3}
  predicate post (k: kont 'a) (arg result: int) = match k with
   | Kid -> let x = arg in x = result
   | Kleft r k' -> let hl = arg in post k' (1 + max hl (height r)) result
   | Kright hl k' -> let hr = arg in post k' (1 + max hl hr) result
   end
\end{why3}
Como pós-condição do construtor \of{Kid} utilizamos a pós-condição trivial
\of{result = x}. Esta fórmula representa a pós-condição mais forte desta função,
podendo ser automaticamente inferida. Utilizando a ferramenta \why\ sobre este
programa, todas as obrigações de prova geradas são automaticamente provadas por
demonstradores SMT.

Um aspecto importante a ressalvar é de que a forma como desfuncionalizámos este
programa e a sua especificação, em particular a forma de construir o predicado
\of{post} e o contrato da função \of{apply}, pode ser mecanizada. Podemos assim
imaginar uma ferramenta que recebe um programa de ordem superior anotado, que o
desfuncionaliza e finalmente o envia à ferramenta \why{}.

\paragraph{Terminação.}
Um aspecto importante que não foi abordado neste exemplo é o da prova de
terminação. Nada nos impede de provar que o programa desfuncionalizado termina,
fornecendo medidas de decréscimo adequadas. Para isso, introduzimos funções
lógicas que contam o número de nós de uma árvore e dos construtores do tipo
\of{kont 'a}. Tais medidas são bastante subtis e requerem um pouco de
imaginação:
\begin{why3}
  function var_tree (t: tree 'a) : int = match t with
    | Empty -> 1
    | Node l _ r -> 3 + var_tree l + var_tree r
    end

  function var_kont (k: kont 'a) : int = match k with
    | Kid -> 0
    | Kleft r k -> 2 + var_tree r + var_kont k
    | Kright _ k -> 1 + var_kont k
    end
\end{why3}
Podemos provar que efectivamente estas funções podem ser utilizadas como medidas
de decréscimo, visto que são limitadas inferiormente:
\begin{why3}
  lemma var_tree_nonneg: forall t: tree 'a. var_tree t >= 0

  lemma var_kont_k_nonneg: forall k: kont 'a. var_kont k >= 0
\end{why3}
Resta-nos juntar as anotações de terminação adequadas ao nosso programa:
\begin{why3}
  let rec height (t: tree 'a) (k: kont 'a) : int
    variant { var_tree t + var_kont k }
    ...
  with apply (k: kont 'a) (arg: int) : int
    variant { var_kont k }
    ...
\end{why3}
Todas as obrigações de prova geradas respeitantes à terminação são também
provadas automaticamente.

Seria interessante ter um mecanismo para fornecer uma especificação sobre a
terminação de um programa de ordem superior e traduzi-la automaticamente, tal
como para o predicado \of{post}.


\subsection{Nombre d'éléments distincts dans un arbre}
\label{sec:nombre-delem-diff}


O próximo exemplo apresenta um programa com o propósito de calcular o número de
elementos distintos de uma árvore binária, com complexidade linear no número de
nós da árvore. Como anteriormente, adoptamos um estilo CPS para evitar qualquer
problema de \textit{stack overflow}. Este programa difere do da secção anterior
pela presença de \emph{efeitos}. Utilizamos um conjunto mutável (uma referência
para um conjunto finito) a fim de guardar os elementos já encontrados. No final
do programa devolvemos o cardinal deste conjunto, obtendo assim o número de
elementos distintos de uma árvore. Segue-se a implementação em \why\ :
\begin{why3}
  let n_distinct_elements (t: tree 'a) : int =
    let h = ref empty in
    let rec distinct_elements_loop (t: tree 'a) (k: unit -> unit) : unit =
      match t with
      | Empty -> k ()
      | Node l x r ->
         h := add x !h;
         distinct_elements_loop l (fun () ->
         distinct_elements_loop r (fun () -> k ()))
      end
    in
    distinct_elements_loop t (fun x -> x);
    cardinal !h
\end{why3}
As operações sobre conjuntos utilizadas neste programa provêm da biblioteca
standard do \why{}. As continuações presentes neste programa têm uma utilização
semelhante às que podemos encontrar no programa da
Sec.~\ref{sec:hauteur-dun-arbre}. A sub-árvore esquerda é tratada pela chamada
\of{distinct_elements_loop l (fun () -> ...)}; a chamada
\of{distinct_elements_loop r (fun () -> ...)} é utilizada para tratar a
sub-árvore direita; para assegurar que de facto devolvemos o resultado correcto,
a primeira chamada a \of{distinct_elements_loop} no interior de
\of{distinct_elements} é feita com a função identidade.


Estando escrito em estilo CPS, este programa combina a utilização de ordem
superior com efeitos. Devemos, por isso, considerar a noção de estado do
programa na sua especificação. Para tal, modificamos a representação das funções
ao nível da lógica, segundo a tese de
J.~Kanig~\cite{kanig10these}. Introduzimos, em primeiro lugar, um novo tipo
\of{state} e estendemos o tipo das funções na lógica como se segue:
\[
  \begin{array}{rl}
    \mathtt{pre} \: f : & \tau_1 \rightarrow\mathtt{state}\rightarrow\mathtt{prop} \\
    \mathtt{post} \: f : &
      \tau_1\rightarrow\mathtt{state}\rightarrow\mathtt{state}\rightarrow\tau_2
                           \rightarrow \mathtt{prop}
  \end{array}
\]
O argumento extra da pré-condição corresponde ao estado no momento da chamada da
função. Os dois argumentos extra da pós-condição representam, respectivamente, o
estado \emph{antes} e \emph{após} a execução da função. A natureza do tipo
\of{state} dependerá das funções consideradas. Para este exemplo, o estado é
simplesmente \of{set 'a}.


Voltemos ao código da função \of{n_distinct_elements}.
Para especificar \of{n_distinct_elements} e \of{distinct_elements_loop}
introduzimos, em primeiro lugar, uma função lógica \of{set_of_tree}. Esta função
calcula a união do conjunto de elementos de uma árvore com um conjunto~\of{s}
dado, que desempenha aqui o papel de acumulador:
\begin{why3}
  function set_of_tree (t: tree 'a) (s: set 'a) : set 'a = match t with
    | Empty -> s
    | Node l x r -> set_of_tree r (set_of_tree l (add x s))
    end
\end{why3}
%
%
Para calcular o conjunto de elementos de uma árvore, basta applicar
\of{set_of_tree} ao conjunto vazio como segundo argumento. Damos à função
\of{n_distinct_elements} o seguinte contrato:
\begin{why3}
  let n_distinct_elements (t: tree 'a) : int
    ensures { result = cardinal (set_of_tree t empty) }
\end{why3}
Especificamos \of{distinct_elements_loop} e as continuações utilizadas de forma
semelhante:
\begin{why3}
  let rec distinct_elements_loop (t: tree 'a) (k: unit -> unit) : unit
    ensures { post k () (set_of_tree t (old !h)) !h () }
  = match t with
    | Empty -> k ()
    | Node l x r ->
       h := add x !h;
       distinct_elements_loop l (fun () -> ensures { post k () (set_of_tree r (old !h)) !h () }
       distinct_elements_loop r (fun () -> ensures { post k () (old !h) !h () }
         k ()))
    end
\end{why3}
A pós-condição de \of{distinct_elements_loop} deve ser alvo de uma explicação
detalhada. Dado que utilizamos a continução~\of{k} no interior de
\of{distinct_elements_loop}, a pós-condição desta função depende da pós-condição
de~\of{k}. É assim necessário caracterizar o estado do programa quando
chamamos~\of{k} e o estado após a sua execução. Este último é o estado obtido
após a execução, por inteiro, da função \of{distinct_elements_loop},
representado pelo valor contido na referência~\of{h}. O estado inicial é mais
subtil. Recordemos que no momento de aplicar~\of{k} teremos já percorrido toda a
árvore~\of{t}. Assim, o estado a partir do qual chamamos a função~\of{k} é o
conjunto de todos os elementos de~\of{t} reunidos com o valor contido em~\of{h}
no momento da chamada a \of{distinct_elements_loop}. Podemos recuperar este
valor através da etiqueta \of{old} do \why{}, que nos permite aceder ao valor
contido numa referência no momento de entrada de uma função.

A especificação das continuações utilizadas no interior das chamadas recursivas
a \of{distinct_elements_loop} segue o raciocínio que acabamos de descrever. As
pós-condições destas duas abstracções dependem, igualmente, da pós-condição
de~\of{k}. Na chamada a \of{distinct_elements_loop l}, especificamos na
pós-condição da sua continuação que o estado com o qual aplicaremos a
continuação é \of{set_of_tree r (old !h)}. Aqui, \of{old !h} representa o estado
antes de se aplicar a continuação, isto é, após se ter percorrido a
sub-árvore~\of{l}. Como esta continuação é utilizada para percorrer toda a
sub-árvore direita, no momento de se aplicar~\of{k} a referência~\of{h} contem,
assim, o conjunto dos elementos distintos de~\of{l} e~\of{x}, que juntámos
inicialmente. Enfim, para a função \of{distinct_elements_loop r}, a sua
continuação nada mais faz que aplicar~\of{k} e por isso esta aplicação é feita
com o estado inicial~\of{old !h}, exactamente o estado antes da chamada
a~\of{k}.

Para desfuncionalizar este programa e traduzir a sua especificação, começamos
por introduzir o tipo algébrico que representa as continuações:
\begin{why3}
  type kont 'a = Kid | Kleft (tree 'a) (kont 'a) | Kright (kont 'a)
\end{why3}
Os únicos efeitos produzidos neste exemplo são o acesso (para escrita e leitura)
à referência~\of{h}. Assim, podemos definir o tipo \of{state} como sendo o tipo
dos valores guardados em~\of{h}:
\begin{why3}
  type state 'a = set 'a
\end{why3}
Finalmente, introduzimos o predicado \of{post}:
\begin{why3}
  predicate post (k: kont 'a) (arg: unit) (old cur: state 'a) (result: unit)
  = match k with
    | Kid -> let () = arg in old == cur
    | Kleft r k -> let () = arg in post k () (set_of_tree r old) cur result
    | Kright k -> let () = arg in post k () old cur result
    end
\end{why3}
Para o caso da função identidade, este predicado especifica que o estado à saída
da função é o mesmo que à entrada. O símbolo \of{(==)} representa a igualdade
extensional entre dois conjuntos. Utilizando esta definição de \of{post},
podemos gerar a função \of{apply} com a sua especificação, como se segue:
\begin{why3}
  let n_distinct_elements (t: tree 'a) : int
    ensures { result = cardinal (set_of_tree t empty) }
  = let h = ref empty in
    let rec apply (k: kont 'a) (arg: unit) : unit
      ensures { post k arg (old !h) !h () }
    = match k with
      | Kid -> let x = arg in x
      | Kleft r k -> let _ = arg in distinct_elements_loop r (Kright k)
      | Kright k -> let _ = arg in apply k arg
      end
    with distinct_elements_loop (t: tree 'a) (k: kont 'a) : unit
      ensures { post k () (set_of_tree t (old !h)) !h () }
    = match t with
      | Empty -> apply k ()
      | Node l x r ->
         h := add x !h;
         distinct_elements_loop l (Kleft r k)
      end
    in
    distinct_elements_loop t Kid;
    cardinal !h
\end{why3}
Uma vez processado pelo \why{}, todas as obrigações de prova geradas para este
programa são automaticamente descartadas. A terminação deste programa poderia
ser igualmente provada, utilizando argumentos de terminação idênticos aos que
foram utilizados para o exemplo anterior. De facto, as medidas de decréscimo
introduzidas para o programa da secção anterior poderiam ser também utilizadas
para este programa, tomando em conta as pequenas diferenças entre os dois tipo
\of{kont}.


\subsection{Interprète à petits pas}
\label{sec:interpr-petits-pas}
O último exemplo que apresentamos é o de um interpretador \textit{small-step}
para uma mini linguagem de expressões aritméticas. Trata-se de uma linguagem
limitada a constantes literais e subtracções:
\begin{why3}
  type exp = Const int | Sub exp exp
\end{why3}
A fim de equipar esta linguagem com uma noção de avaliação, definimos a seguinte
função lógica \of{eval}:
\begin{why3}
  function eval (e: exp) : int = match e with
    | Const n -> n
    | Sub e1 e2 -> (eval e1) - (eval e2)
    end
\end{why3}
Esta função representa a semântica natural (\textit{big-step}) da nossa
linguagem.

Para definir uma relação de redução, começamos por definir a relação
$\xrightarrow \epsilon$, que corresponde a uma redução à cabeça da
expressão. Para esta linguagem existe uma só regra de redução à cabeça:
\[
\mathtt{Sub} \; (\mathtt{Const}\: \mathtt{v_1})\; (\mathtt{Const}\: \mathtt{v_2})  \quad
\xrightarrow \epsilon \quad \mathtt{Const}\: (\mathtt{v_1 - v_2})
\]
Para traduzir $\xrightarrow \epsilon$ em \why\, introduzimos a função
\of{head_reduction}:
\begin{why3}
  let head_reduction (e: exp) : exp = match e with
    | Sub (Const v1) (Const v2) -> Const (v1 - v2)
    | _ -> absurd
    end
\end{why3}
O segundo ramo da filtragem represente o caso em que uma expressão passada como
argumento não é um \textit{redex}. Como desejamos aplicar \of{head_reduction}
unicamente a expressões que possam ser reduzidas à cabeça, utilizamos a
palavra-chave \of{absurd} do \why\ para marcar este ramo como um ponto
inatingível do código. Para provar que este ramo é efectivamente
inatingível\footnote{A construção \of{absurd} exige uma prova de falso.}, é
necessário exigir que o argumento de \of{head_reduction} seja um
\textit{redex}. Para isso, introduzimos o seguinte predicado \of{is_redex}:
\begin{why3}
  predicate is_redex (e: exp) = match e with
    | Sub (Const _) (Const _) -> true
    | _ -> false
    end
\end{why3}
Podemos agora juntar a \of{head_reduction} a pré-condição desejada:
\begin{why3}
  let head_reduction (e: exp) : exp
    requires { is_redex e }
    ...
\end{why3}
O resultado da chamada \of{head_reduction e} é uma expressão~\of{e'} que se
avalia ao mesmo valor que~\of{e}. Juntamos a \of{head_reduction} uma
pós-condição que especifica exactamente este raciocínio:
\begin{why3}
  let head_reduction (e: exp) : exp
    requires { is_redex e }
    ensures  { eval result = eval e }
    ...
\end{why3}
Todas as obrigações de prova geradas para a função \of{head_reduction} são
automaticamente descartadas.

Para reduzir em profundidade, introduzimos agora a regra de inferência
\[
\frac{\mathtt{e}\xrightarrow \epsilon \mathtt{e'}}{C[\mathtt{e}]\rightarrow C[\mathtt{e'}]}
\]
onde $C$ representa um contexto de redução, definido pela seguinte gramática:
\[
  \begin{array}{crl}
    C & ::= & \square \\
      & |   & C[\mathtt{Sub} \; \square \; \mathtt{e}] \\
      & |   & C[\mathtt{Sub} \; (\mathtt{Const} \: \mathtt{v}) \; \square]
  \end{array}
\]
O símbolo $\square$ representa o buraco. Definimos $C[\mathtt{e}]$ como sendo o
contexto $C$ no qual o buraco foi completamente substituído pela
expressão~\of{e}. A esta operação damos o nome de \emph{composition}. O
resultado desta operação é uma nova expressão. A gramática apresentada induz uma
construção de contextos \textit{bottom-up}. A nossa escolha de contextos define,
ainda, uma ordem de avaliação por valor e da esquerda para a direita.

O ponto interessante deste exemplo é a maneira como escolhemos representar os
contextos. Em detrimento de uma representação baseada num tipo algébrico com
três construtores, escolhemos aqui representar um contexto como uma função. Um
contexto~\of{c} é assim uma função que recebe como argumento uma
expressão~\of{e} e devolve a expressão obtida por substituição do buraco
de~\of{c} para uma expressão~\of{e}:
\begin{why3}
  type context = exp -> exp
\end{why3}
O contexto vazio, o buraco, é representado pela função identidade:
\begin{why3}
  let hole = fun x -> x
\end{why3}
Os contextos para a redução sobre o primeiro ou sobre o segundo argumento de
\of{Sub} são representados, respectivamente, como se segue:
\begin{why3}
  let sub_left e2 c = fun e1 -> c (Sub e1 e2)
  let sub_right v c = fun e1 -> c (Sub (Const v) e1)
\end{why3}
Esta maneira de representar os contextos conduz-nos a um código de ordem
superior elegante, mais compacto que um código utilizando um tipo concreto para
os contextos, tal como observaremos em seguida.

Tendo definida a noção de contexto, a próxima etapa na implementação de um
interpretador \textit{small-step} para a nossa linguagem consiste na
implementação de uma função que decompõe uma expressão~\of{e} em um
\textit{redex}~\of{e'} e um contexto~$C$, tais que
$C[\mathtt{e'}]=\mathtt{e}$. Para a nossa linguagem esta decomposição é
única. Tal operação pode ser implementada com base na seguinte função auxiliar
\of{decompose_term}:
\begin{why3}
  let rec decompose_term (e: exp) (c: context) : (context, exp) =
    match e with
    | Const _ -> absurd
    | Sub (Const v1, Const v2) -> (c, e)
    | Sub (Const v, e) -> decompose_term e (fun x -> c (Sub (Const v) x))
    | Sub (e1, e2) -> decompose_term e1 (fun x -> c (Sub x e2))
    end
\end{why3}
Explicamos agora em detalhe o funcionamento da função \of{decompose_term}. Esta
função recebe como argumento um contexto~\of{c} e uma expressão~\of{e} que
desejamos decompor. Para isso, procedemos a uma análise por casos sobre a forma
da expressão~\of{e}. Não podemos decompor um valor, logo utilizamos \of{absurd}
para marcar o primeiro ramo como um ponto inatingível. Se a expressão é da forma
\of{Sub (Const v1) (Const v2)}, trata-se de um \textit{redex}. Encontramos assim
a decomposição que procurávamos, com a função \of{decompose_term} a devolver
este redex e o contexto~\of{c}. Se, pelo contrário, existem em~\of{e}
sub-expressões que possamos ainda reduzir, devemos continuar o processo de
decomposição. Se a primeira sub-expressão de \of{Sub} se encontra já reduzida,
continuamos a decomposição sobre a segunda sub-expressão com um novo contexto de
redução sobre a segunda sub-expressão. Se, por seu lado, a primeira
sub-expressão de \of{Sub} não está ainda reduzida, a decomposição continua com
um contexto de redução sobre a primeira sub-expressão\footnote{O leitor notará
  que o código das funções \of{hole}, \of{sub_left} e \of{sub_right} se encontra
expandido na construção dos contextos.}

Apresentamos em seguida a especificação lógica para a função
\of{decompose_term}. Pretendemos aplicar \of{decompose_term} unicamente a
expressões que não são ainda valores. Para tal, introduzimos uma função lógica
\of{is_value} com o propósito de testar se uma expressão é uma constante:
\begin{why3}
  predicate is_value (e: exp) = match e with
    | Const _ -> true
    | _ -> false
    end
\end{why3}
Munidos deste predicado, podemos agora equipar \of{decompose_term} e as
abstracções utilizadas no seu interior com os devidos contratos:
\begin{why3}
  let rec decompose_term (e: exp) (c: context) : (context, exp)
    requires { not (is_value e) }
    ensures  { let (c', e') = result in
                 is_redex e' && forall res. post c e res -> post c' e' res }
  = match e with
    ...
    | Sub (Const v, e) ->
       decompose_term e (fun x -> ensures { post c (Sub (Const v) x) result } ...)
    | Sub (e1, e2) ->
       decompose_term e1 (fun x -> ensures { post c (Sub x e) result } ...)
    end
\end{why3}
De notar que este programa não apresenta efeitos. Consequentemente, utilizamos
uma projecção \of{post} de dois argumentos, em semelhança ao que fizemos na
Sec.~\ref{sec:hauteur-dun-arbre}. A função \of{decompose_term} deve verificar a
propriedade $C'[\mathtt{e'}] = C[\mathtt{e}]$, $C'$ e $\mathtt{e'}$ sendo o
contexto e a expressão devolvida e $C$ et $\mathtt{e}$ os argumentos. Ora, dado
que escolhemos representar os contextos como funções, esta operação de
composição corresponde precisamente à aplicação de um contexto ao seu
argumento. A fim de especificar esta aplicação, utilizamos a projecção \of{post}
sobre um contexto e a quantificação universal para referir que para qualquer
expressão~\of{res} que verifica \of{post c e res}, então \of{post c' e' res} é
igualmente verificada. Por outro lado, a pós-condição de \of{decompose_term}
assegura (à esquerda do símbolo \of{&&}) que a expressão~\of{e'} devolvida é um
\textit{redex}.

Passamos agora a introduzir a função \of{decompose} e a sua especificação. Esta
função toma como argumento uma expressão e não faz mais que chamar
\of{decompose_term}, dando-lhe como argumento o contexto vazio, aqui
representado pela função identidade:
\begin{why3}
  let decompose (e: exp) : (context, exp)
    requires { not (is_value e) }
    ensures  { let (c', e') = result in is_redex e' && post c' e' e  }
  = decompose_term e (fun x -> x)
\end{why3}
Para avaliar uma expressão~\of{e} em um valor, introduzimos uma função de
iteração que se comporta como o fecho transitivo da relação~$\rightarrow$. O seu
código \why\ é o seguinte:
\begin{why3}
  let rec red (e: exp) : int = match e with
    | Const v -> v
    | _ ->
      let (c, r) = decompose e in
      let     r' = head_reduction r in
      red (c r')
    end
\end{why3}
Se a expressão~\of{e} é da forma \of{Const v}, esta encontra-se já em forma
normal e por isso não pode ser reduzida. Se, por outro lado,~\of{e} não é ainda
uma constante temos então de (1) decompor esta expressão no
\textit{redex}~\of{r} e no contexto~\of{c}; (2) reduzir~\of{r} à cabeça e obter
a expressão~\of{r'}; (3) continuar a iteração com a expressão obtida pela
composição de~\of{c} e~\of{r'}. A composição materializa-se, de uma forma
elegante, como a aplicação de~\of{c} a~\of{r'}. O valor devolvido pela chamada
\of{red e} deve ser o mesmo que aquele que é devolvido por uma avaliação
\textit{big-step}. Equipamos assim \of{red} com o contrato seguinte:
\begin{why3}
  let rec red (e: exp) : int
    ensures { result = eval e }
\end{why3}

Procedemos à desfuncionalização deste exemplo para mostrar a sua correcção
através do \why{}. Começamos por introduzir o tipo \of{context}
desfuncionalizado:
\begin{why3}
  type context = CHole | CApp_left context exp | CApp_right int context
\end{why3}
A partir das abstracções presentes no programa original, podemos gerar a função
\of{apply} conjuntamente com a sua especificação:
\begin{why3}
  let rec apply (c: context) (arg: exp) : exp
    ensures { post c arg result }
  = match c with
    | CHole -> let x = arg in x
    | CApp_left c e -> let x = arg in apply c (Sub x e)
    | CApp_right v c -> let x = arg in apply c (Sub (Const v) x)
    end
\end{why3}
onde o predicado \of{post} pode ser deduzido das pós-condições igualmente
presentes no programa original:
\begin{why3}
  predicate post (c: context) (arg result: exp) = match c with
    | CHole -> let x = arg in x = result
    | CApp_left c e -> let x = arg in post c (Sub x e) result
    | CApp_right v c -> let x = arg in post c (Sub (Const v) x) result
    end
\end{why3}
Para obter o programa desfuncionalizado final, basta, assim como demonstrado
para os exemplos anteriores, substituir todas as aplicações das abstracções por
chamadas a \of{apply} e todas as ocorrências dessas mesmas abstracções pelo
construtor respectivo do tipo \of{context} desfuncionalizado.

Dando ao \why{} o programa desfuncionalizado e a respectiva especificação, tal
como apresentados, seriamos capazes de provar todas as obrigações de prova
geradas, excepto a pós-condição da função \of{red}. Para ajudar os
demonstradores automáticos a descartar esta pós-condição, introduzimos o
seguinte lema auxiliar:
\begin{why3}
  lemma post_eval: forall c arg1 arg2 r1 r2.
    eval arg1 = eval arg2 -> post c arg1 r1 -> post c arg2 r2 ->
    eval r1 = eval r2
\end{why3}
Este lema exprime que para qualquer expressão~\of{arg1} e~\of{arg2} cuja a
avaliação produz o mesmo valor, então as expressões obtidas pela composição de
\of{arg1} e \of{arg2} com o mesmo contexto~\of{c} devem-se avaliar no mesmo
valor. Provamos este resultado por indução sobre~\of{c} e para tal escrevemos um
\emph{lemma function}, isto é, um programa fantasma que termina e que não possui
efeitos observáveis, cujo contrato será automaticamente traduzido para o
enunciado apresentado acima:
\begin{why3}
  let rec lemma post_eval (c: context) (arg1 arg2 r1 r2: exp)
    requires { eval arg1 = eval arg2 }
    requires { post c arg1 r1 && post c arg2 r2 }
    ensures  { eval r1 = eval r2 }
    variant  { c }
  = match c with
    | CHole -> ()
    | CApp_left c e  -> post_eval c (Sub arg1 e) (Sub arg2 e) r1 r2
    | CApp_right n c ->
        post_eval c (Sub (Const n) arg1) (Sub (Const n) arg2) r1 r2
    end
\end{why3}
As chamadas recursivas deste programa simulam a aplicação da hipótese de
indução. Com este lemma auxiliar, a pós-condição de~\of{red} é provada
automaticamente por demonstradores SMT. O enunciado deste lema quantifica
universalmente sobre valores do tipo \of{context}. Escrever tal enunciado
directamente sobre o programa de ordem superior poderia conduzir a uma
contradição, visto que não nos é possível afirmar tal resultado para qualquer
função do tipo $\mathtt{exp}\rightarrow\mathtt{exp}$. Ainda assim, seria
interessante ter um mecanismo que nos permitisse escrever este género de
enunciados no código original (restringindo as funções abstractas aceites),
traduzindo-os automaticamente para o código desfuncionalizado.

É interessante notar que o tipo dos contextos desfuncionalizado corresponde a um
\emph{zipper}~\cite{zipper} para expressões da nossa linguagem. Esta forma de
representar os contextos permite-nos implementar uma função de avaliação
eficaz. Esta função é normalmente designada como
\emph{refocusing}~\cite{DBLP:journals/tcs/DanvyN01}. Utilizando contextos como
funções, as nossas operações de decomposição e composição apresentam uma
complexidade linear no pior caso, o que confere à função \of{red} um custo
potencialmente quadrático sobre o tamanho de uma expressão.

\section{Discussão e perspectivas}
\label{sec:conclusion}

Neste artigo explorámos a utilização da desfuncionalização como uma técnica de
prova para programas de ordem superior contendo potencialmente efeitos. O que
propomos é que um programa de ordem superior seja directamente anotado, e em
seguida desfuncionalizar este programa e a sua especificação para um programa de
primeira ordem. Uma ferramenta de verificação dedutiva existente pode ser então
utilizada para demonstrar a correcção deste programa. Esta abordagem permite-nos
utilizar ferramentas de verificação dedutiva de programas de primeira ordem
existentes, sem necessidade de estender estas mesmas ferramentas com suporte
para ordem superior (o que implicaria, muito provavelmente, uma mudança profunda
no \textit{core} destas ferramentas, nomeadamente ao nível da sua linguagem de
programação, lógica subjacente e cálculo de obrigações de prova). Ilustrámos a
nossa proposta através de três exemplos desfuncionalizados (manualmente). Os
programas de primeira ordem obtidos foram automaticamente verificados no sistema
\why{}. A utilização da desfuncionalização num contexto de prova é, tanto quanto
sabemos, nova.

A desfuncionalização é geralmente considerada uma técnica global de
transformação de programas, isto é, é necessário conhecer todas as funções
utilizadas como valores de primeira classe num programa no momento de aplicar
esta transformação. Tal cenário implica que não poderemos ambicionar provar
todos os programas de ordem superior através da desfuncionalização. Ainda assim,
o método proposto aplica-se com sucesso a programas escritos em estilo CPS,
visto tratarem-se de programas para os quais conhecemos todas as abstracções
utilizadas em funções de ordem superior.

\paragraph{Perspectivas.} O trabalho apresentado neste artigo mantém-se, para
já, exploratório. Utilizámos a desfuncionalização para provar diversos exemplos
de programas de ordem superior e os resultados encorajam-nos a continuar a
exploração desta metodologia. O nosso objectivo é melhorar e formalizar a nossa
utilização da desfuncionalização sobre programas de ordem superior. O primeiro
passo será definir formalmente a classe de programas sobre os quais esta técnica
pode ser aplicada, a fim de compreender se a desfuncionalização pode ser
utilizada como uma técnica robusta de prova de programas de ordem superior com
efeitos.

A função~\of{apply}, gerada durante a desfuncionalização, simula a aplicação de
uma função (representada como um valor de um tipo algébrico obtido por
desfuncionalização) ao seu argumento. A função \of{apply} procede por análise de
casos sobre cada função presente no programa original, onde cada ramo
corresponde à definição da abstracção correspondente. Assim, cada um destes
ramos deve devolver um valor do mesmo tipo, o que é apenas verdade quando todas
as abstracções do programa têm o mesmo tipo $\tau_1\rightarrow\tau_2$. Para
resolver este problema, Pottier e Gauthier~\cite{pottier-gauthier-hosc}
propuseram a utilização de \emph{tipos algébricos generalizados} (do inglês
\emph{generalized algebraic data types}, GADT) para desfuncionalizar um programa
contendo abstracções de diferentes tipos. Tal solução interessa-nos e por isso
pretendemos estudar possíveis mecanismos para estender a linguagem e lógica do
sistema \why\ com suporte para os GADT. Por agora, introduzimos diferentes
funções \of{apply} caso o programa inicial apresente abstracções de tipos
diferentes.

A fim de apresentar garantias fortes sobre a fiabilidade de um programa, a sua
prova de terminação desempenha um papel fundamental. Neste artigo mencionámos
como poderia ser demonstrado que um programa desfuncionalizado termina, mas
sempre através de uma intervenção manual sobre o programa gerado. Seria, por
isso, desejável podermos especificar directamente a terminação no programa de
ordem superior com medidas de decréscimo apropriadas, que seriam depois
traduzidas durante a desfuncionalização.

Os iteradores são exemplos de funções de ordem superior frequentemente
utilizadas para enumerar os elementos de uma estrutura de dados. Em trabalhos
anteriores~\cite{pereira16nfm,pereira16jfla}, propusemos técnicas para provar a
correcção de programas que implementam tais iteradores, assim como programas que
os utilizam. Para tal, iniciámos o desenvolvimento de uma ferramenta que recebe
como \textit{input} programas de ordem superior anotados e que se serve do \why\
para gerar obrigações de prova. Pretendemos estender esta ferramenta com
suporte para a desfuncionalização e aumentar, assim, o conjunto de programas de
ordem superior aceites pela ferramenta.

Finalmente, para melhor avaliar a sua utilidade, desejamos aplicar a prova por
desfuncionalização a um caso de estudo mais complexo. Um bom candidato é o
algoritmo de Koda-Ruskey~\cite{KodaRuskey93} para a enumeração de todos os
ideais de um conjunto parcialmente ordenado. É nossa intenção partir da
implementação de ordem superior existente deste algoritmo~\cite{Filliatre01a}, e
provar a sua correcção através da nossa metodologia.








\section*{Agradecimentos}

Agradeço a Lucas Baudin, Richard Bonichon, Martin Clochard e Léon Gondelman
pelos comentários e sugestões durante a preparação deste artigo. Agradeço
profundamente ao Jean-Christophe Filliâtre. Os seus conselhos, re-leituras
atentas e correcções foram cruciais para a conclusão deste trabalho.

\bibliographystyle{plain}
\bibliography{local_bib}

\begin{thebibliography}{10}

\bibitem{DBLP:conf/ppdp/AgerBDM03}
Mads~Sig Ager, Dariusz Biernacki, Olivier Danvy, and Jan Midtgaard.
\newblock A functional correspondence between evaluators and abstract machines.
\newblock In {\em Proceedings of the 5th International {ACM} {SIGPLAN}
  Conference on Principles and Practice of Declarative Programming, 27-29
  August 2003, Uppsala, Sweden}, pages 8--19. {ACM}, 2003.

\bibitem{chargueraud11icfp}
Arthur Chargu{\'e}raud.
\newblock Characteristic formulae for the verification of imperative programs.
\newblock In Manuel M.~T. Chakravarty, Zhenjiang Hu, and Olivier Danvy,
  editors, {\em Proceeding of the 16th ACM SIGPLAN international conference on
  Functional Programming (ICFP)}, pages 418--430, Tokyo, Japan, September 2011.
  ACM.

\bibitem{danvy09scp}
Olivier Danvy and Kevin Millikin.
\newblock Refunctionalization at work.
\newblock {\em Science of Computer Programming}, 74(8):534--549, 2009.
\newblock Special Issue on Mathematics of Program Construction (MPC 2006).

\bibitem{danvy01ppdp}
Olivier Danvy and Lasse~R. Nielsen.
\newblock Defunctionalization at work.
\newblock In {\em Proceedings of the 3rd ACM SIGPLAN International Conference
  on Principles and Practice of Declarative Programming}, PPDP '01, pages
  162--174. ACM Press, 2001.

\bibitem{DBLP:journals/tcs/DanvyN01}
Olivier Danvy and Lasse~R. Nielsen.
\newblock Syntactic theories in practice.
\newblock {\em Electr. Notes Theor. Comput. Sci.}, 59(4):358--374, 2001.

\bibitem{Filliatre01a}
Jean-Christophe Filli\^atre.
\newblock La sup\'eriorit\'e de l'ordre sup\'erieur.
\newblock In {\em Journ\'ees Francophones des Langages Applicatifs}, pages
  15--26, Anglet, France, January 2002.

\bibitem{filliatre13esop}
Jean-Christophe Filli\^atre and Andrei Paskevich.
\newblock Why3 --- where programs meet provers.
\newblock In Matthias Felleisen and Philippa Gardner, editors, {\em Proceedings
  of the 22nd European Symposium on Programming}, volume 7792 of {\em Lecture
  Notes in Computer Science}, pages 125--128. Springer, March 2013.

\bibitem{pereira16jfla}
Jean-Christophe Filli\^atre and M\'ario Pereira.
\newblock It\'erer avec confiance.
\newblock In {\em Vingt-septi\`emes Journ\'ees Francophones des Langages
  Applicatifs}, Saint-Malo, France, January 2016.

\bibitem{pereira16nfm}
Jean-Christophe Filli\^atre and M\'ario Pereira.
\newblock A modular way to reason about iteration.
\newblock In Sanjai Rayadurgam and Oksana Tkachuk, editors, {\em 8th NASA
  Formal Methods Symposium}, volume 9690 of {\em Lecture Notes in Computer
  Science}, Minneapolis, MN, USA, June 2016. Springer.

\bibitem{zipper}
G\'erard Huet.
\newblock {The Zipper}.
\newblock {\em Journal of Functional Programming}, 7(5):549--554, September
  1997.

\bibitem{kanig10these}
Johannes Kanig.
\newblock {\em Sp\'ecification et preuve de programmes d'ordre sup\'erieur}.
\newblock Th{\`e}se de doctorat, Universit{\'e} Paris-Sud, 2010.

\bibitem{KanigFilliatre09wml}
Johannes Kanig and Jean-Christophe Filli\^atre.
\newblock {Who: A Verifier for Effectful Higher-order Programs}.
\newblock In {\em ACM SIGPLAN Workshop on ML}, Edinburgh, Scotland, UK, August
  2009.

\bibitem{KodaRuskey93}
Yasunori Koda and Frank Ruskey.
\newblock {A Gray Code for the Ideals of a Forest Poset}.
\newblock {\em Journal of Algorithms}, (15):324--340, 1993.

\bibitem{nanevski08icfp}
Aleksandar Nanevski, Greg Morrisett, Avi Shinnar, Paul Govereau, and Lars
  Birkedal.
\newblock {Ynot}: Reasoning with the awkward squad.
\newblock In {\em Proceedings of ICFP'08}, 2008.

\bibitem{Plotkin75}
Gordon~D. Plotkin.
\newblock Call-by-name, call-by-value and the lambda-calculus.
\newblock {\em Theor. Comput. Sci.}, 1(2):125--159, 1975.

\bibitem{pottier-gauthier-hosc}
François Pottier and Nadji Gauthier.
\newblock Polymorphic typed defunctionalization and concretization.
\newblock {\em Higher-Order and Symbolic Computation}, 19:125--162, March 2006.

\bibitem{regis-gianas-pottier-08}
Yann R\'egis-Gianas and Fran\c{c}ois Pottier.
\newblock A {Hoare} logic for call-by-value functional programs.
\newblock In {\em Proceedings of the Ninth International Conference on
  Mathematics of Program Construction (MPC'08)}, pages 305--335, 2008.

\bibitem{reynolds-98a}
John~C. Reynolds.
\newblock Definitional interpreters for higher-order programming languages.
\newblock {\em Higher-Order and Symbolic Computation}, 11(4):363--397, December
  1998.

\end{thebibliography}

\end{document}